\documentclass[letter]{ptptex}

\usepackage{graphicx}
\usepackage{wrapft}

\notypesetlogo                       

\title{
Baryonic ${}^3 P_2$-Dominant Superfluidity under Combined \\
Pion Condensation with $\Delta$-Mixing
}

\author{
Ryozo {\sc Tamagaki}\footnote{E-mail: tama-ktn@nike.eonet.ne.jp} 
and Tatsuyuki {\sc Takatsuka}$^{*,}$
\footnote{E-mail: takatuka@iwate-u.ac.jp}
}

\inst{
Kamitakano Maeda-Cho 26-5, Kyoto 606-0097, Japan \\
$^{*}$Faculty of Humanities and Social Sciences, Iwate University \\
Morioka 020-8550, Japan
}

\abst{
We study baryonic superfluidity under the combined pion condensation with $\Delta$-mixing, where both condensates of the neutral 
and charged pions take place in neutron stars. Outline of formulation treating such superfluidity is presented 
and energy gap calculations show the possibility that critical temperatures are higher than about $1\times 10^{9}$ K.
}

\begin{document}

\maketitle

\newcommand{\gsim}{>\kern-12pt\lower5pt\hbox{$\displaystyle\sim$}}
\newcommand{\lsim}{<\kern-12pt\lower5pt\hbox{$\displaystyle\sim$}}
\newcommand{\mbf}{\boldsymbol}


(1) {\it Introduction}\hspace{5mm}
Recent experiments on the spin-isospin excitation mode of nuclei\cite{SS98} have shown that condensations of the neutral pions ($\pi^{0}$) and 
the charged pions ($\pi^{c}$) likely set in at almost same density not higher than about two times the nuclear density 
($\rho_0=0.17 {\rm fm}^{-3}$) with the assist of the $\Delta$-degrees of freedom.\cite{Tatsumi03}  
This view on pion condensation (PC) differs from the previous one that the charged PC firstly occurs at about $2\rho_0$ and the neutral PC appears 
at the higher density, e.g., $\rho\simeq (3-5)\rho_0$.\cite{MT87} The combined PC means that both condensates of $\pi^{0}$ and $\pi^{c}$ 
coexist in neutron stars (NSs). Such relatively low threshold density of the combined PC signifies its existence in most of NSs except for 
those with very small masses. 
   
This paper is the first to study the baryonic superfluid (SF) under the combined PC with the $\Delta$-degrees of freedom.
Here we adopt a model of the combined PC, studied previouly,\cite{TamTam78,MT87} that both condensates contribute to energy gain 
without interference. The baryon system under this combined PC takes the specific structure which we have called the 
{\it alternating layer spin} (ALS) structure.\cite{SuppCh3,SuppCh5} The SF in such a combined PC consists of the pair with the spin-triplet, 
two-dimensional (2D) $P$-wave and 2D total angular momentum $m_J=\pm 2$, that is, the pair with the streched 2D angular momentum.  
This is because the pair can utilize the attractive effects from the $LS$ interaction, needed for the occurrence of the SF 
at $\rho \gsim \rho_0$, as shown in previous studies by the present authors.\cite{SuppCh2,SuppCh4,TT9799} 
As this SF contains the ${}^3 P_2$-pair as a dominant one, we call it the ${}^3 P_2$-dominant SF. 
This paper presents outline of a formulation to treat this SF and briefly reports calculated results of energy gaps, equivalently 
critical temperatures. Full description will be reported in a succeeding paper.

(2) {\it Quasi-baryons ($\tilde{n}$ and $\tilde{p}$) and quasi-particles ($\eta$) in the combined PC}\hspace{5mm}
The ground state of the system without the pairing correlation, denoted as $|\Phi_0\rangle$, is represented by the direct product 
of the fermion ground state  $|\Phi_F\rangle$ and the boson ground state representing the combined PC: 
$|\Phi_0\rangle = |\Phi_F\rangle \times |\Phi_B\rangle$.\cite{SuppCh3} The $|\Phi_F\rangle$ is described by the Fock state of the quasi-particles 
denoted by the $\eta$ particles: $|\Phi_F\rangle =\prod_{\beta}^{({\rm occ})}\eta_{\beta}^{\dagger}|0 \rangle$,
which tends to the neutron matter without PC. The $\eta$ particle signifying the single-particle eigen-mode in the combined PC 
is given by a superposition of a quasi-neutron ($\tilde{n}$) and a quasi-proton ($\tilde{p}$) constructed in the following way.

Due to the action of the condensed $\pi^{0}$ field with the momentum ${\mbf k}_0$, the quasi-neutron $\bar{n}$ (the quasi proton $\bar{p}$) 
is formed as a superposition of the neutron $n$ (the proton $p$) and the $\Delta^{0}$ ($\Delta^{+}$) with a same charge and spin 
component.\cite{SuppCh3,SuppCh5} Due to the action of the condensed $\pi^{c}$ field with 
the momentum ${\mbf k}_c$, a quasi-neutron $\tilde{n}$ (a quasi-proton $\tilde{p}$) is formed as a superposition of the $\bar{n}$ ($\bar{p}$) 
with the spin$=\pm1/2$ and the $\Delta^{-}$ ($\Delta^{++}$) of the spin state  
$|\Delta_{(\pm 1/2)} \rangle\equiv \pm \hat{k}_{\pm}|\Delta_{\mp 1/2}\rangle/2 
\mp \sqrt{3} \hat{k}_{\mp}|\Delta_{\pm 3/2}\rangle/2$. \cite{MT87} Here $\hat{k}_{\pm}\equiv (k_{cx}\pm ik_{cy})/k_c={\rm exp}(i\sigma \theta_c)$ 
for $\sigma=\pm$, where $\theta_c$ is the angle between ${\mbf k}_c$ and the $x$-axis, since we take the angular-momentum quantization 
axis along ${\mbf k}_0$ in the $z$ direction, and ${\mbf k}_c$ to be perpendicular to ${\mbf k}_0$,

The $\eta$-particle is described by a superposition of $\tilde{n}$ and $\tilde{p}$ with opposite spin components, 
because the nucleon-$\pi^{c}$ interaction flips the spin and isospin. Tranformation of the quasi-baryon operators 
to the $\eta$ operators and its inverse are written as follows: \cite{TamTam78}
\begin{subequations}
\label{eq:1}
\begin{equation}
\eta_{\beta}=u_{\beta}^{*}\tilde{n}_{\beta}-v_{\beta}^{*}\tilde{p}_{\beta -},\;\;\; 
\zeta_{\beta}=v_{\beta}\tilde{n}_{\beta}+u_{\beta}\tilde{p}_{\beta -}, 
\label{eq:1a}
\end{equation}
\begin{equation}
\tilde{n}_{\beta}=u_{\beta}\eta_{\beta}+v_{\beta}^{*}\zeta_{\beta},\;\;\;
\tilde{p}_{\beta -}=-v_{\beta}\eta_{\beta}+u_{\beta}^{*}\zeta_{\beta}.
\end{equation}
\end{subequations}
The $\zeta$-states orthogonal to the $\eta$-states are empty because of their energy much higher than the $\eta$'s. 
The suffices are given as $\beta \equiv \{ {\mbf q}_{\perp},q_z^{(\pi)},\sigma \}$ and  
$\beta- \equiv \{({\mbf q}_{\perp}-{\mbf k}_{c}),q_z^{(\pi)},-\sigma \}$, where $({\mbf q}_{\perp},q_z)$ denotes the momentum, 
$\sigma/2$ is the spin component and $\pi=+(-)$ means the ``$\ell$-parity" of the Bloch-orbital state constructed from the localized 
wave functions around layer centers of $\ell=$even (odd) numbers. 
The $u,\;v$ factors satisfy the relation $|u_{\beta}|^{2}+|v_{\beta}|^{2}=1$. We have the relation, $\pi=\pm$ for $\sigma=\pm 1$, 
according to our specification of the ALS state.  The coefficients are expicitly given as
$u_{\beta} =u_{{\mbf q}_{\perp},q_z^{(\pi)}}\simeq u$ and  
$v_{\beta} =-i{\rm e}^{i\sigma \theta_c}v_{{\mbf q}_{\perp},q_z^{(\pi)}}\simeq -i {\rm e}^{i\sigma \theta_c}v$.
In the last lines the approximation to neglect weak momentum dependence of the $u,\; v$ factors is used, and then 
$u^{2}+v^{2}=1$. In treating the $\Delta$-degrees of freedom in the SF problems, we basically follow previous papers.\cite{TT9799}

(3) {\it Quasi-baryon pair operators and pairing interaction}\hspace{5mm}
Pairing correlation under the combined PC works through the side of the Fermi cylinder, namely for the 
$({\mbf q}_{\perp},{\mbf -q}_{\perp})$ pair. Because of the band gap just outside the upper and lower Fermi surfaces, 
the pairing correlation does not work with respect to the momentum $q_z$. 

We consider the quasi-baryon pair operators with the 2D streched angular momentum, namely, with the spin $m_S=\sigma/2+\sigma/2=\sigma$, 
the spatial $m_{L}=\sigma$, and the total $m_J= m_S+m_{L}=2\sigma=\pm 2$ for $\sigma=\pm 1$.     
By this choice we have $m_S m_L>0$ which gives the attraction from the $LS$ potential because its radial part in the triplet-odd state 
is negative.

There are two independet groups of the quasi-baryon pairs, the one with $m_J=+2$ and the one with $m_J=-2$, because $m_J$ is the good 
quantum number. The aspects of the $m_J=-2$ group are the same with those of the $m_J=+2$ group except that all the components of 
the 2D angular momenta are reversed.  Hereafter we treat only the $m_J=+2$ group fixed by $\sigma=+1$ for the derivation of the 
gap equation and so forth, although the total energy shift due to the pairing correlation is the sum of the contributions from the two groups.  

We define three quasi-baryon pair operators with $\Lambda\equiv \{|m_S|=1,\;|m_L|=1\}$ and $m_J=+2$, 
for the $\tilde{n}\tilde{n}$-pair, the $\tilde{p}\tilde{p}$-pair and 
the symmetrized $(\tilde{n}\tilde{p})$-pair, as follows:
\begin{subequations}
\label{eq:1}
\begin{equation}  
B^{(\tilde{n}\tilde{n})\dagger}_{\Lambda m_J}(q_{\perp},q_z) \equiv \frac{1}{\sqrt{2}}\int\;d\varphi_q\; \frac{{\rm e}^{i\sigma\varphi_q}}
{\sqrt{2\pi}}\;\tilde{n}^{\dagger}_{{\mbf q}_{\perp},q_z^{(+)},\sigma}\;\tilde{n}^{\dagger}_{-{\mbf q}_{\perp},-q_z^{(+)},\sigma}, 
\label{eq:1a}
\end{equation}
\begin{eqnarray}
B^{(\tilde{p}\tilde{p})\dagger}_{\Lambda m_J}(q_{\perp},q_z)& \equiv & \frac{1}{\sqrt{2}}\int\;d\varphi_q \;
\frac{{\rm e}^{i\sigma\varphi_q}}{\sqrt{2\pi}}\;\tilde{p}^{\dagger}_{({\mbf q}_{\perp}-{\mbf k}_c),q_z^{(-)},\sigma}
 \; \tilde{p}^{\dagger}_{(-{\mbf q}_{\perp}-{\mbf k}_c),-q_z^{(-)},\sigma}, \\
B^{(\tilde{n}\tilde{p})\dagger}_{\Lambda m_J}(q_{\perp},q_z)& \equiv &  \frac{1}{\sqrt{2}}\int\;d\varphi_q\; 
\frac{{\rm e}^{i\sigma\varphi_q}}{\sqrt{2\pi}}
\;\Big\{ \tilde{n}^{\dagger}_{{\mbf q}_{\perp},q_z^{(+)},\sigma}\;\tilde{p}^{\dagger}_{(-{\mbf q}_{\perp}-{\mbf k}_c),-q_z^{(-)},\sigma} \nonumber \\
& &+ \tilde{p}^{\dagger}_{({\mbf q}_{\perp}-{\mbf k}_c),q_z^{(-)},\sigma}
\;\tilde{n}^{\dagger}_{-{\mbf q}_{\perp},-q_z^{(+)},\sigma} \Big\}.
\end{eqnarray}
\end{subequations} 
In these $B^{\dagger}$, $\int \;d\varphi_q\; {\rm e}^{i\sigma\varphi_q}$ plays a role to make the projection to the 2D  
orbital angular momentum $m_L=\sigma$, and thus leads to $m_J=2\sigma$. Owing to this projection operator, the quasi-baryon operators 
are symmetric for $q_z \rightarrow -q_z$, and we can restrict $q_z$ to $q_z>0$ in these pair operators.  

Discriminating these quasi-baryon pair operators by ${\cal P}$, we define the most favorable pairing interaction in the combined PC as follows: 
\begin{eqnarray}
H_{\rm BB-pair}^{(\Lambda m_J=+2)}=\sum_{q'_{\perp}}\sum_{q_z'}\sum_{q_{\perp}}\sum_{q_z}\sum_{{\cal P}'}\sum_{{\cal P}}
\;\langle q'_{\perp},q_z',{\cal P}'|V_{\rm BB}^{(\Lambda m_J)}|q_{\perp},q_z,{\cal P}\rangle \nonumber \\
\times B_{\Lambda m_J}^{({\cal P}') \dagger}(q'_{\perp},q_z') B_{\Lambda m_J}^{({\cal P})}(q_{\perp},q_z). 
\end{eqnarray}
By extracting $\eta^{\dagger}$ from $B^{({\cal P})\dagger}(q_{\perp},q_z)$ using Eq. (1b) as     
\begin{equation}
\tilde{n}^{\dagger}_{{\mbf q}_{\perp},q_z^{(+)},\sigma}\rightarrow u\;\eta^{\dagger}_{{\mbf q}_{\perp},q_z^{(+)},\sigma}, \;\;\; 
\tilde{p}^{\dagger}_{({\mbf q}_{\perp}-{\mbf k}_c),q_z^{(-)},\sigma}\rightarrow -i{\rm e}^{i\sigma\theta_c}v\;
\eta^{\dagger}_{{\mbf q}_{\perp},q_z^{(-)},-\sigma},
\end{equation}
we have the $\eta$-pair operators $B^{(\Pi)\dagger}(q_{\perp},q_z)$ expressed by $\eta^{\dagger}$ as follows:\\
$B^{(\tilde{n}\tilde{n})\dagger}_{\Lambda m_J}(q_{\perp},q_z) \equiv u^{2}\;B_{\Lambda m_J}^{(\Pi=+1)\;\dagger}(q_{\perp},q_z)$,  
$B^{(\tilde{p}\tilde{p})\dagger}_{\Lambda m_J}(q_{\perp},q_z) \equiv -v^{2}\;B_{\Lambda m_J}^{(\Pi=-1)\;\dagger}(q_{\perp},q_z)$, and \\ 
$B^{(\tilde{n}\tilde{p})\dagger}_{\Lambda m_J}(q_{\perp},q_z) \equiv (-iuv)\;B_{\Lambda m_J}^{(\Pi=0)\;\dagger}(q_{\perp},q_z)$.

Denoting the coefficients as $C_{\Pi=+1}=u^{2},\;\; C_{\Pi=-1}=-v^{2},\;C_{\Pi=0}=iuv$
we have the pairing interaction Hamiltonian written in terms of $\eta$ and $\eta^{\dagger}$: 
\begin{eqnarray} 
 H_{\rm BB-pair}^{(\Lambda m_J=+2)}  =  \sum_{q'_{\perp}}\sum_{q_z'}\sum_{q_{\perp}}\sum_{q_z}\sum_{\Pi'}\sum_{\Pi} 
\langle q_{\perp}',q_z',{\cal P}'|C_{\Pi'}^{*}V_{\rm BB}^{(\Lambda m_J)}C_{\Pi}|q_{\perp},q_z,{\cal P} \rangle \nonumber \\
 \times  B_{\Lambda m_J}^{(\Pi')\dagger}(q_{\perp}',q_z')\;B_{\Lambda m_J}^{(\Pi)}(q_{\perp},q_z),
\end{eqnarray}
where the restriction for $q_z>0$ is not necessary for the matrix elements of the pairing interaction. 
It is to be noted that the matrix elements are taken with respect to the states of the quasi-baryon pair, not to those of 
the $\eta$ pair, despite of the one to one correspondence between ${\cal P}$ and $\Pi$.

(4) {\it Bogoliubov transformation and coupled gap equation}\hspace{5mm} 
In treating a coupled pairing of three channels with respect to $\Pi$, we apply the Bogoliubov transformation following the method  
previously developed in the study of the ${}^3 P_2$-pairing, referred to as T70.\cite{T70} Namely, the superfluid groud state $|\Psi_0\rangle$ is obtained 
by operating a unitary tansformation ${\rm e}^{iS}$ on the nonsuperfluid ground state $|\Phi_F\rangle $:
\begin{equation}
|\Psi_0\rangle = {\rm e}^{iS} |\Phi_F\rangle =
{\rm e}^{iS_{\Lambda m_J=+2}}\times {\rm e}^{iS_{\Lambda m_J=-2}}|\Phi_F\rangle .
\end{equation}

We first express $iS_{\Lambda m_J=+2}$ part using $B^{(\Pi) \dagger}$ and $B^{(\Pi)}$ and then rewrite it in terms of $\eta$ and $\eta^{\dagger}$, 
using ${\mbf q}\equiv ({\mbf q}_{\perp},q_z)$,  as follows:
\begin{equation}
iS_{\Lambda m_J} = \frac{1}{2}\sum_{\mbf q}\sum_{\pi_1\pi_2}\Big\{ 
\theta_{\Lambda m_J}({\mbf q},\pi_1,\pi_2)\eta^{\dagger}_{{\mbf q},\pi_1\sigma}\eta^{\dagger}_{-{\mbf q},\pi_2\sigma}
-\theta^{*}_{\Lambda m_J}({\mbf q},\pi_1,\pi_2)\eta_{-{\mbf q},\pi_2\sigma}\eta_{{\mbf q},\pi_1\sigma} \Big\},
\end{equation}
The suffix $\sigma$ can be suppressed because $\sigma=+1$ is fixed for $m_J=+2$, and the transformation is carried out with respect to $\pi$. 
As for the formal procedure, we can proceed in pararell with T70, treating $\pi$ in place of $\sigma$. 
The $\theta$ function has the symmetric and parity-odd property: 
$\theta({\mbf q},\pi_1,\pi_2)=\theta({\mbf q},\pi_2,\pi_1)$ and $\theta(-{\mbf q},\pi_1,\pi_2)=-\theta({\mbf q},\pi_1,\pi_2)$.
We write the operators $\eta$ and  $\eta^{\dagger}$ columnwise as
\begin{eqnarray}
& & \vec{ \eta}_{\mbf q}= \left(
\begin{array}{c}
\eta_{{\mbf q},+}  \\
\eta_{{\mbf q},-}   
\end{array}
\right) , 
\hspace{1cm} \vec{\eta}_{\mbf q}\,^{\dagger}= \left(  
\begin{array}{cc}
\eta_{{\mbf q},+}^{\dagger}    \\
\eta_{{\mbf q},-}^{\dagger}    
\end{array}
\right),
\end{eqnarray}
and  similarly write the transformed operators $\alpha$ and $\alpha^{\dagger}$, which we abbreviate as SFQP for the superfluid quasi-particle.   
Then, denoting  $\theta({\mbf q},\pi_1,\pi_2)$ by the $(2\times 2)$ matrix ${\mbf \Theta}({\mbf q})$, we have the Bogoliubov transformation:
\begin{subequations}
\label{eq:1}
\begin{equation}
\vec{\alpha}_{{\mbf q}} \equiv {\rm e}^{iS_{\Lambda m_J}}\vec{\eta}_{{\mbf q}}{\rm e}^{-iS_{\Lambda m_J}}
= {\mbf U}({\mbf q})\vec{\eta}_{{\mbf q}}-{\mbf V}({\mbf q})\vec{\eta}_{-{\mbf q}}\,^{\dagger}, 
\label{eq:1a}
\end{equation}
\begin{equation}
\vec{\alpha}_{{\mbf q}}\,^{\dagger} \equiv {\rm e}^{iS_{\Lambda m_J}}\vec{\eta}_{{\mbf q}}\,^{\dagger}{\rm e}^{-iS_{\Lambda m_J}}
= \tilde{{\mbf U}}({\mbf q})\vec{\eta}_{{\mbf q}}\,^{\dagger}-{\mbf V}^{\dagger}({\mbf q})\vec{\eta}_{-{\mbf q}}. 
\end{equation}
\end{subequations}
${\mbf U}$ and ${\mbf V}$ are the $(2\times 2)$ matrices with respect to $\pi$, which are given by 
\begin{subequations}
\label{eq:1}
\begin{equation}
{\mbf U}({\mbf q})  \equiv  {\rm cos}{\mbf \Theta}'({\mbf q}), 
\label{eq:1a}
\end{equation}
\begin{equation}
{\mbf V}({\mbf q})  \equiv  {\mbf \Theta}'({\mbf q})^{-1}{\rm sin}{\mbf \Theta}'({\mbf q}){\mbf \Theta}({\mbf q}),
\end{equation}
\end{subequations}

where ${\mbf \Theta}'({\mbf q})$ is the Hermite matrix determined by the relation
${\mbf \Theta}'({\mbf q})^{2}={\mbf \Theta} ({\mbf q}){\mbf \Theta} ({\mbf q})^{\dagger}$.

We rewrite the pairing interaction Hamiltonian using the inverse Bogoliubov transformation in terms of $\alpha$ and $\alpha^{\dagger}$ 
and lead it into the reduced form consisting of the constant term without $\alpha^{\dagger}$ and $\alpha$, the single SFQP term 
with $\alpha^{\dagger}\alpha$, and the dangerous term involving $\alpha\alpha$ and $\alpha^{\dagger}\alpha^{\dagger}$. 
From the vanishing of the dangerous term, we reach a coupled gap equation among three gap functions $\Delta_{\Pi}(q_{\perp},q_z)$ 
of $\Pi=+1,\;-1,\;0$:
\begin{equation}
\Delta_{\Pi}(q_{\perp},q_z)=-\frac{1}{2}\sum_{\Pi'}\sum_{q'_{\perp}}\sum_{q'_z}
\langle q_{\perp},q_z,{\cal P}|C_{\Pi}^{*}V_{\rm BB}^{(\Lambda m_J)}C_{\Pi'}|q'_{\perp},q_z',{\cal P}'\rangle
K_{\Pi'}\frac{\Delta_{\Pi'}(q'_{\perp},q_z')}{E(q'_{\perp},q_z')},
\end{equation}
where $E(q_{\perp},q_z)=\sqrt{\tilde{\epsilon}_{\eta}^{2}(q_{\perp},q_z)+D^{2}(q_{\perp},q_z)}$, 
$\tilde{\epsilon}_{\eta}(q_{\perp},q_z)\simeq \hbar^{2}q_{\perp}^{2}/2M_n m_{\eta}^{*}$ ($m_{\eta}^{*}$ being an effective mass 
parameter of the $\eta$-particle), $K_{\Pi =+1,-1,0}=1,1,2$ and the squared energy gap $D^{2}=\sum_{\Pi}K_{\Pi}\Delta_{\Pi}^{2}/2\pi$. 
In getting this familiar form of the SFQP energy, we have taken an approximation that a squared gap matrix (${\mbf D}^{2}$) 
is replaced by an appropriate c-number form for which we choose its trace. 

(5) {\it Estimate of energy gap and critical temparature}\hspace{5mm} 
For the NS matter in the combined PC, the $\tilde{n}$ is the dominant component, even though the $\tilde{p}$ is generated by the 
charged PC, and in the pairing problem the $\Pi=+1$-channel corresponding to the $\tilde{n}\tilde{n}$-pair contributes most importantly. 
Generally, small components play a role to enhance the main component in the coupled gap equation. Typical examples can be seen in the 
tensor-force coupling in the ${}^3 P_2 +{}^3 F_2$-pairing\cite{Takatsuka72} and the ${}^3 S_1 +{}^3 D_1$-pairing.\cite{SuppCh2} 
Here the $\Pi=-1$ and $\Pi=0$ channels enhance the gap function of the dominant $\Pi=+1$ channel through the coupling terms. 
Thus we can estimate the lower limit of energy gaps by making calculations in the $\Pi=+1$ channel only. 

\begin{figure}[b]
 \parbox{\halftext}{
    \centerline{\includegraphics[width=5cm,height=5cm]{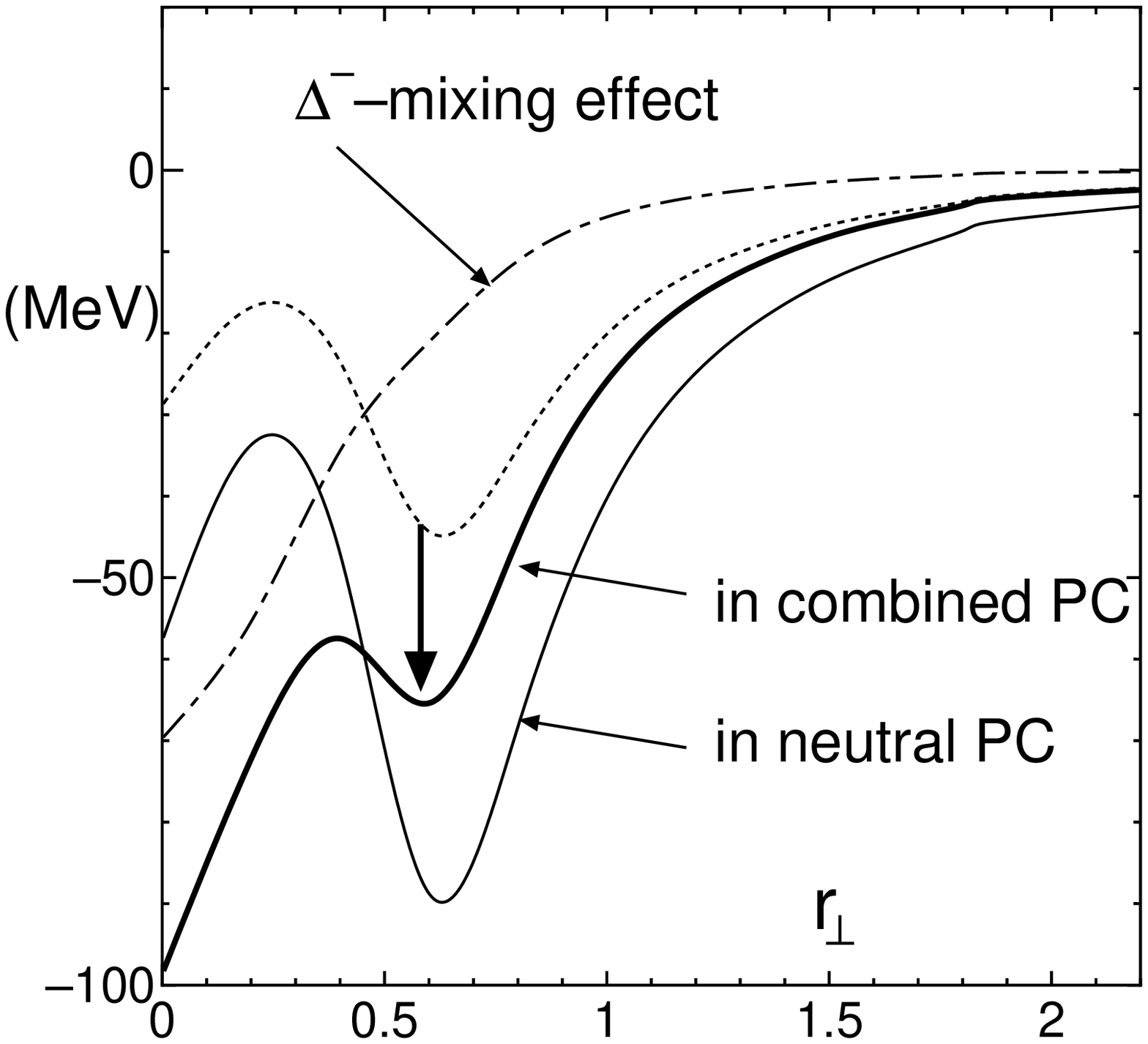}}
     \caption{The thick solid curve shows the effective 2D pairing potential of the main $\tilde{n}\tilde{n}$ channel in the combined PC 
obtained for OPEG-B potential and at $\rho=3.5\rho_0$. The corresponding potential in the neutral PC 
(the one multiplied by the attenuation factor=0.500) is shown by the solid thin (dotted) curve. The dash-dotted curve shows the attraction 
 due to the $\Delta^{-}$-mixing}
\label{fig:1}}
 \hfill
 \parbox{\halftext}{
 \centerline{\includegraphics[width=6cm,height=6cm]{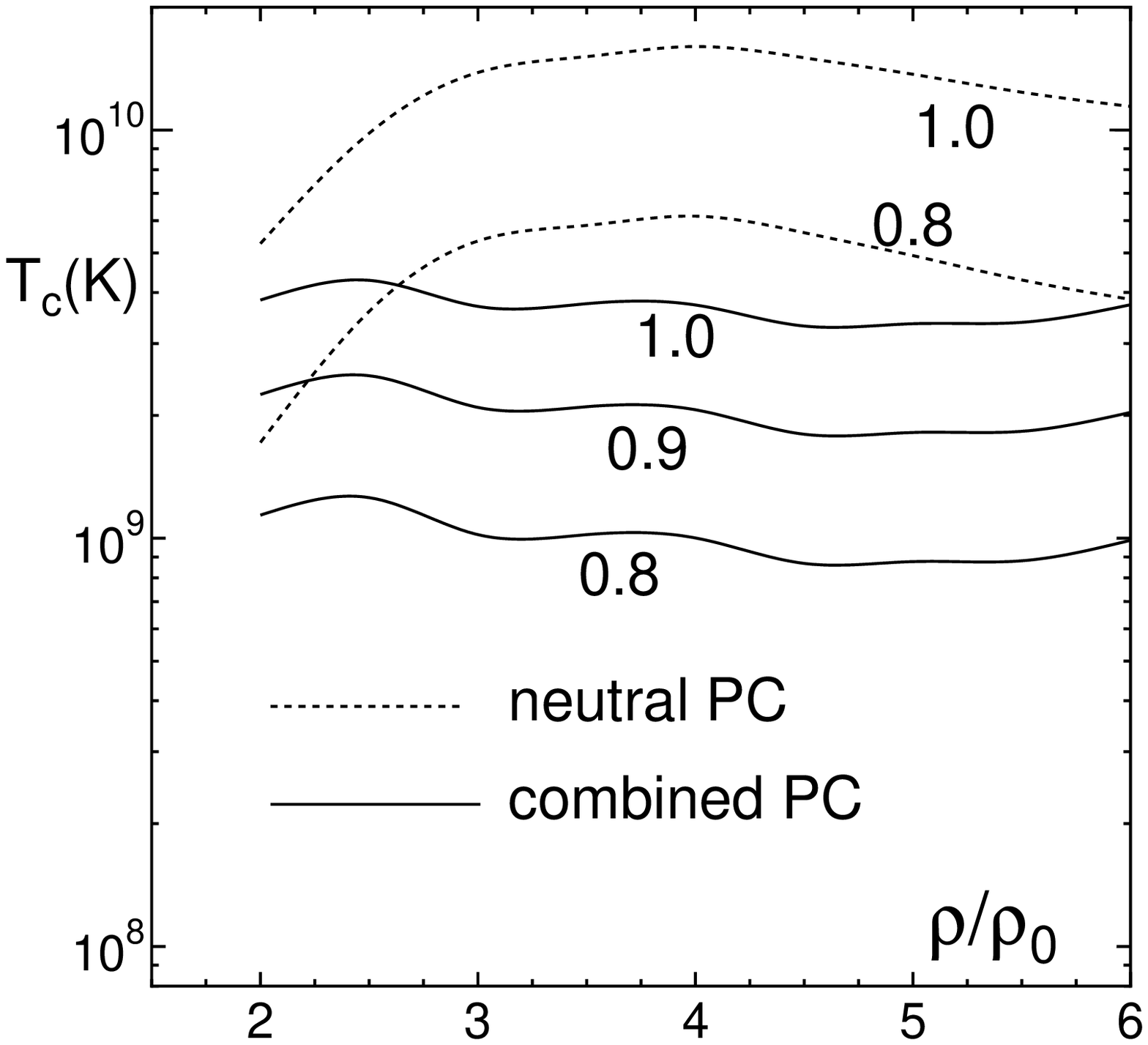}}
     \caption{The solid (dotted) curves show the density dependence of SF critical temperatures $T_c$ in the combined PC (the neutral PC) 
for OPEG-B potential. The numbers attached to the curves are the effective mass parameters $m_{\eta}^{*}$.} 
\label{fig:2}}
\end{figure}  

The matrix elements in the main channel (${\cal P}=\tilde{n}\tilde{n}$) is approximately given by the following 2D radial integral, 
 after properly managing the normalization factors, as follows: With use of the cylindrical coordinate $(r_{\perp},\phi,z)$,
\begin{equation}
\langle  q_{\perp},q_z,{\cal P}|C_{1}^{2} V_{\rm BB}^{(\Lambda m_J)}|q_{\perp}',q_z',{\cal P} \rangle
\simeq \frac{(1-v^{2})^{2}}{(1+y_1^{2})^{2}} \int _0^{\infty}dr_{\perp}r_{\perp}J_1(q_{\perp}r_{\perp})
\tilde{V}_{{\cal P}}^{(\Lambda m_J)}(r_{\perp}) J_1(q_{\perp}'r_{\perp}),
\end{equation}
where $v^{2}$ is the $\tilde{p}$-mixing probability, $y_1^{2}$ the $\Delta^{-}$-mixing probability and $J_1(q_{\perp}r_{\perp})$ 
the Bessel function.  Here we take an approximation suitable for the 
well-developed ALS structure, because this structure grows just beyond the on-set density of the neutral PC.\cite{SuppCh3} 
Then the $q_z$ dependence becomes irrelevant. 
$\tilde{V}_{{\cal P}}^{(\Lambda m_J)}(r_{\perp})$ is the 2D effective potential obtained after the integation over $z$ and $\phi$, where 
we suppress the $q_{\perp}$ dependence although include it in solving the gap equation.  
In the single channel we have $D(q_{\perp})=\Delta_{+1}(q_{\perp})/\sqrt{2\pi}$.

Energy gap calculations are done by adopting the baryon-baryon potential working in the $(N+\Delta)$ space extended from $NN$ OPEG-B 
potential\cite{mopeg} and by using the PC parameters for the strong PC case given in refs. \citen{MT87,SuppCh8} but slightly modified 
so as to have smooth density dependence. Fig.1 shows an example of the 2D effective potential calculated at $\rho=3.5\rho_0$. 
Attenuation of the pairing attraction caused due to the charged PC is severe as $(1-v^{2})^{2}/(1+y_1^{2})^{2} 
\simeq 0.74\rightarrow 0.50\rightarrow 0.63$, for such density variation as $v^{2}\simeq 0.10\rightarrow 0.24\rightarrow 0.17$ 
and $y_1^{2}\simeq 0.045\rightarrow 0.072\rightarrow 0.042$ at $\rho=(2\rightarrow 4 \rightarrow 6)\rho_0$. 
But the $\Delta^{-}$-mixing brings about the short-range attraction shown by the dot-dashed curve in Fig. 1, and the pairing attraction 
is recovered from the attenuated one considerably as illustrated by the thick arrow.   
As a result, the reduction of the pairing attraction from that in the neutral PC is gentle. 

Superfluid critical temperatures $T_c$ obtained from energy gap calculations, \linebreak $k_B T_c\simeq 0.57 D(q_{\perp F})$ 
($q_{\perp F}$ being the Fermi momentum), are shown in Fig. 2 at densities $\rho=(2-6)\rho_0$ and for several values of 
the effective mass parameters of the $\eta$-particle $m_{\eta}^{*}$. Since the $m_{\eta}^{*}$ in the combined PC are about (0.8-0.9), 
we have $T_c \gsim 1 \times 10^{9}$ K for this interaction. 
The result is meanigful, because such $T_c$ efficiently suppress too rapid cooling due to the pion direct Urca process 
taking place in most of NSs, except for those with very small masses. 
This point is to be considered in the studies of NS cooling.\cite{UNTMT94,Tsuruta8698,YKGH01,Tsuruta03,YP04} 
To proceed further, we need detailed studies on the matrix elements of the channel coupling 
and solutions of the coupled gap equation, which will be reported elsewhere.

The authors thank S.~Tsuruta and T.~Tatsumi for their cooperative discussions.

\end{document}